\documentclass[conference]{IEEEtran}
\IEEEoverridecommandlockouts

\usepackage{cite}
\usepackage{amsmath,amssymb,amsfonts}
\usepackage{algorithmic}
\usepackage{graphicx}
\usepackage{textcomp}
\usepackage{xcolor}
\usepackage{subfigure}
\usepackage{caption, subcaption}
\usepackage{amsmath,graphicx}
\usepackage{textcomp}
\usepackage{amsmath,amssymb,amsfonts}
\usepackage{mathtools}
\usepackage{amsmath,graphicx, multirow, hyperref}

\newcommand{\cognospeak}{CognoSpeak}
\newcommand{\bart}{BART}
\newcommand{\albert}{AlBERT}
\newcommand{\bert}{BERT}
\newcommand{\roberta}{RoBERTa}
\newcommand{\distilbert}{DistilBERT}
\newcommand{\egemaps}{eGeMAPS}
\newcommand{\compare}{ComParE}
\newcommand{\etal}{\textit{et al.}}

\def\BibTeX{{\rm B\kern-.05em{\sc i\kern-.025em b}\kern-.08em
    T\kern-.1667em\lower.7ex\hbox{E}\kern-.125emX}}
\begin{document}

This paper has been accepted for publication at IEEE SSCI 2025. \\

The copyright belongs to IEEE. 

\title{\cognospeak{}: an automatic, remote assessment of early cognitive decline in real-world conversational speech\\
}

\author{\IEEEauthorblockN{Madhurananda Pahar$^1$\IEEEauthorrefmark{1},
Fuxiang Tao$^1$\IEEEauthorrefmark{1},
Bahman Mirheidari$^1$\IEEEauthorrefmark{1},
Nathan Pevy$^1$\IEEEauthorrefmark{2},
Rebecca Bright$^2$\IEEEauthorrefmark{3},\\ 
Swapnil Gadgil$^2$\IEEEauthorrefmark{3}, Lise Sproson$^3$\IEEEauthorrefmark{4},
Dorota Braun$^4$\IEEEauthorrefmark{1},
Caitlin Illingworth$^4$\IEEEauthorrefmark{1},\\
Daniel Blackburn$^4$\IEEEauthorrefmark{1},
Heidi Christensen$^1$\IEEEauthorrefmark{1}
}
\vspace{5pt}
\IEEEauthorblockA{$^1$School of Computer Science, University of Sheffield, Sheffield, S1 4DP, UK\\
$^2$Therapy Box, London, UK\\
$^3$NIHR Devices for Dignity HTC, Sheffield Teaching Hospitals NHS Foundation Trust, Sheffield, S10 2JF, UK\\
\vspace{5pt}
$^4$Sheffield Institute for Translational Neuroscience (SITraN), University of Sheffield, Sheffield, S10 2HQ, UK\\
Email: \IEEEauthorrefmark{1}\{m.pahar, f.tao, b.mirheidari, d.a.braun, c.illingworth, d.blackburn, heidi.christensen\}@sheffield.ac.uk,\\ \IEEEauthorrefmark{2}nathanpevy@gmail.com,
\IEEEauthorrefmark{3}\{rbright, sgadgil\}@therapy-box.co.uk, \IEEEauthorrefmark{4}lise.sproson@nihr.ac.uk
}
}

\maketitle

\begin{abstract}
    The early signs of cognitive decline are often noticeable in conversational speech, and identifying those signs is crucial in dealing with later and more serious stages of neurodegenerative diseases. Clinical detection is costly and \mbox{time-consuming} and although there has been recent progress in the automatic detection of speech-based cues, those systems are trained on relatively small databases, lacking detailed metadata and demographic information. This paper presents \cognospeak{} and its associated data collection efforts. 
    \cognospeak{} asks memory-probing long and short-term questions and administers standard cognitive tasks such as verbal and semantic fluency and picture description using a virtual agent on a mobile or web platform. 
    In addition, it collects multimodal data such as audio and video along with a rich set of metadata from primary and secondary care, memory clinics and remote settings like people's homes. 
    Here, we present results from 126 subjects whose audio was manually transcribed. 
    Several classic classifiers, as well as large language model-based classifiers, have been investigated and evaluated across the different types of prompts. 
    We demonstrate a high level of performance; in particular, we achieved an $F_1$-score of 0.873 using a \distilbert{} model to discriminate people with cognitive impairment (dementia and people with mild cognitive impairment (MCI)) from healthy volunteers using the memory responses, fluency tasks and cookie theft picture description. \cognospeak{} is an automatic, remote, low-cost, repeatable, \mbox{non-invasive} and less stressful alternative to existing clinical cognitive assessments. 
\end{abstract}


\vspace{-3pt}


\section{Introduction}
\label{sec:intro}

Struggles with memory and cognition can stem from a variety of causes, including fatigue, stress, and illness, and often worsen with age. 
If such issues persist beyond a few months, they might indicate mild cognitive impairment (MCI) \cite{davis2018estimating}, a condition linked to problems with memory, learning, reasoning, attention, conversation, language, and loss of interest or motivation. 
Those experiencing MCI often describe their condition as `brain fog' as it affects their ability to think clearly \cite{rosenberg2013association}. 
Approximately half of those diagnosed with MCI eventually develop dementia, a progressive condition caused by brain-damaging diseases including Alzheimer's (AD) \cite{prestia2013prediction}. 
Other forms of dementia, such as vascular and frontotemporal dementia, often affect behaviour and cognitive abilities such as language, perceptual and executive functions \cite{knopman2003essentials, thabtah2020correlation}. 
Both MCI and dementia are the initial stages of cognitive decline, 
and lack a cure \cite{hendrie1998epidemiology}.

The benefits of detecting early signs of dementia include timely treatment to delay the later stages. However, this can be challenging as it requires thorough investigations by neurological experts and maintaining historical records through cognitive assessment by the health service providers \cite{shi2023speech}. 
Existing clinical methods for diagnosis are brain magnetic resonance imaging measurement (MRI), scale testing and cerebrospinal fluid analysis \cite{yang2022deep}, which are expensive, \mbox{time-consuming}, unpleasant to the participants and laborious; which is why these methods 
are disadvantageous for \mbox{large-scale} cognitive decline screening \cite{mckhann2011diagnosis}. 
This is also why 75\% of individuals with early cognitive decline do not get treated at all \cite{gauthier2021world}, making it one of the most under-diagnosed conditions for the global ageing population. 
There is, therefore, an increased necessity for remote, smart technologies to support healthcare services to deliver timely and accurate diagnosis \cite{gauthier2021world}. 


\begin{figure}[t]
  \centering 
  \includegraphics[width=\linewidth]{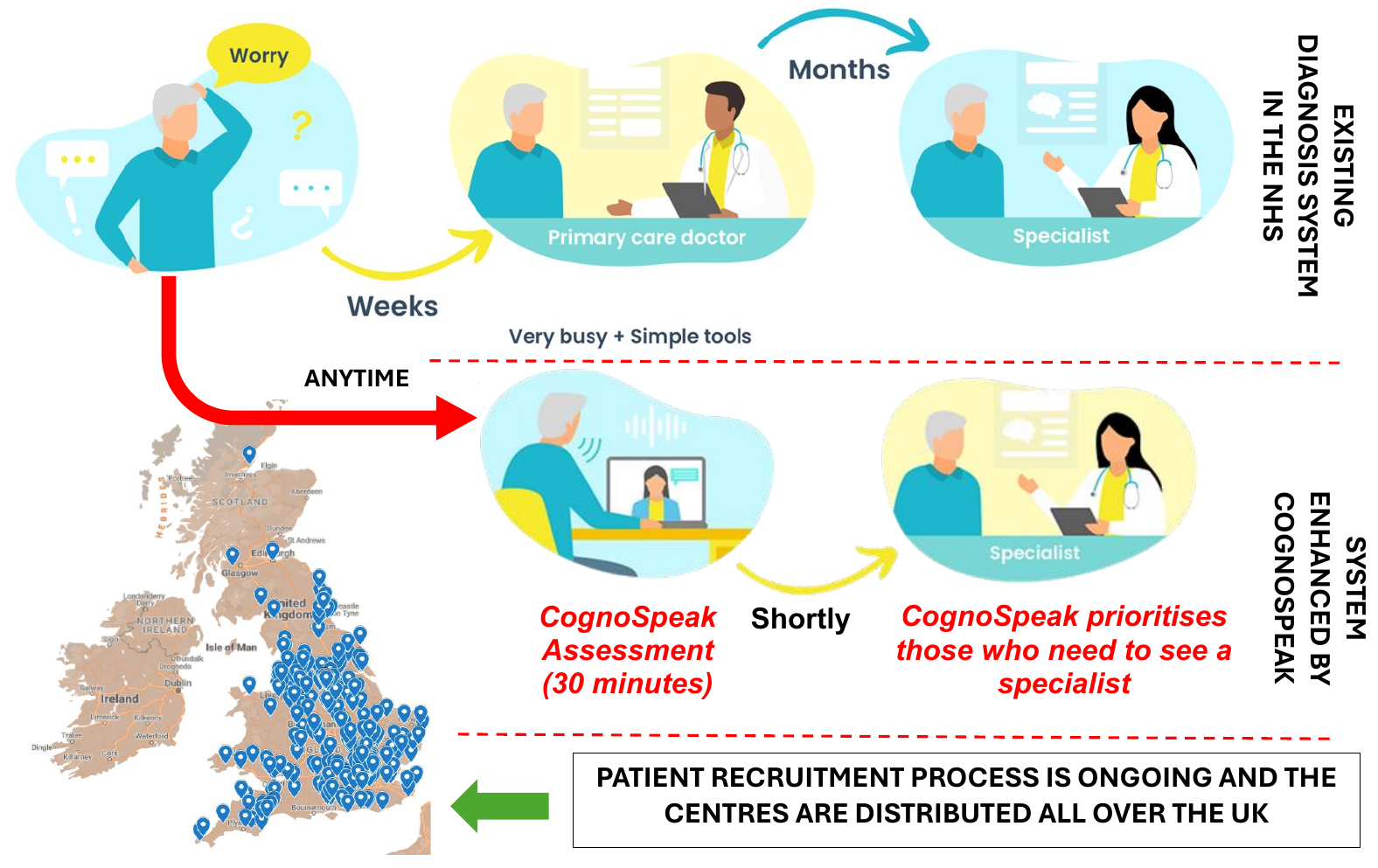} 
  \caption{\textbf{\cognospeak{}} will expedite the stratification process for those who are concerned about their cognitive health as it is capable of accurately distinguishing between those who show signs of cognitive decline and therefore need referring to a more specialist assessment and those who have other causes of their memory problems such as depression or anxiety. Currently, data collection is ongoing in various parts of the UK, encompassing a wide range of accents and demographics. 
  Participants are recruited nationwide through primary and secondary care,  various websites such as \href{https://www.joindementiaresearch.nihr.ac.uk/}{Join Dementia Research} and social media channels, including a number of memory clinics for the study.
  }
  \label{fig:cognospeak_novelty}
\end{figure}

\begin{figure*}[t]
  \centering 
  \includegraphics[width=\linewidth]{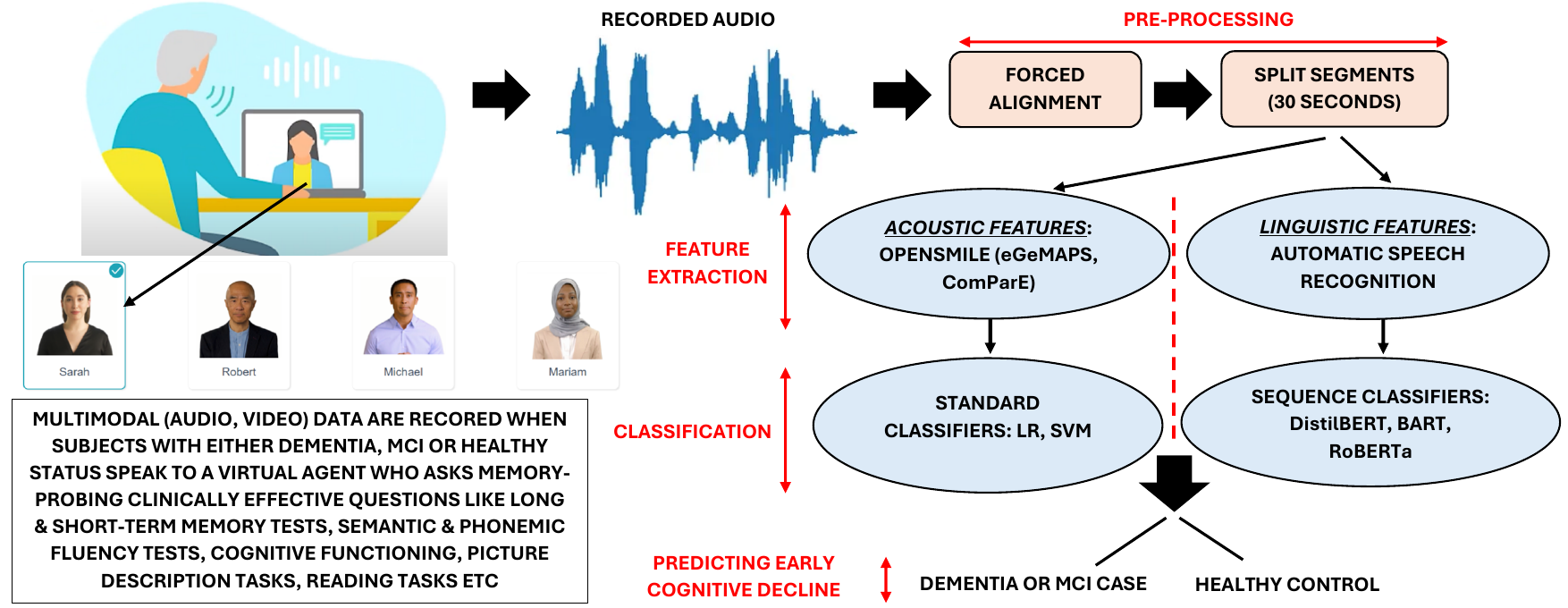} 
  \caption{\textbf{\cognospeak{}} collects real-world audio and video recordings when a virtual agent prompts the subject for a diverse range of clinically proven effective tasks on multiple platforms such as mobile and web. Four avatars (2 male and 2 female) from diverse ethnic groups were used as the virtual agent. The recorded audio is then pre-processed and both acoustic and linguistic features are extracted to train and evaluate classifiers such as standard classifiers and sequence classifiers (foundation models). Finally, those with either dementia or mild cognitive impairment (MCI) are distinguished from healthy.}
  \label{fig:cognospeak}
\end{figure*}

\cognospeak{} is designed to accelerate the identification of early cognitive decline, thus saving time and financial resources for health service providers across the world, such as the National Health Service (NHS) in the UK (Figure \ref{fig:cognospeak_novelty}). 
Using a virtual agent on a laptop or tablet, \cognospeak{} engages the subject in a conversation which is cognitively demanding on multiple domains involving memory, language and attention and carefully analyses the spoken answers to detect early signs of dementia as described in Figure \ref{fig:cognospeak}. 
The virtual agent administers a diverse range of tasks such as long and \mbox{short-term} memory tests, semantic and phonemic fluency tests, picture description tasks, and reading tasks. Gold-standard clinical processes have diagnosed the subjects' cognitive status into three categories which are dementia, MCI, and healthy control (HC). Subjects are invited for \mbox{follow-up} assessments to enable longitudinal studies. In addition, the subjects undergo multiple standard cognitive and mood assessments, which are further described in Section~\ref{section:expsetup}, to allow for a range of detection and regression experimental analyses.
\cognospeak{} is currently used to collect data in primary and secondary care in the UK as well as recruiting nationally through websites like 
\href{https://www.joindementiaresearch.nihr.ac.uk/}{Join Dementia Research}. 
This remote, large-scale data collection, either at patients' homes, at clinics or in community centres, allows us to capture real-world conversational speech. 
To the best of our knowledge, this is the largest collection of this type of data. 


This paper presents the first results from the initial data collection phase of audio recordings from 126 subjects. 
Next, in Section \ref{sec:previouswork}, similar work carried out in the past and their limitations are discussed, followed by 
Section \ref{section:expsetup}, where
the system and the initially collected data are described. 
Section \ref{section:expsetup} also describes
the experimental setup of feature extraction, classification and evaluation. 
The results are summarised in Section \ref{sec:results} and Section \ref{sec:discussion} discusses them.
Finally, Section \ref{sec:conclusion} concludes that \cognospeak{}, an automatic, low-cost and remote assessment tool, is a promising viable means to provide a diagnostic aid for the early detection and tracking of signs of cognitive decline for healthcare providers. 
Python scripts used in this study are shared via our GitHub repository \cite{madhu_pahar_2024_14515541}. 

\begin{table*}[t!]
    \centering
    \setlength{\tabcolsep}{7pt} 
    \renewcommand\arraystretch{1.2}
    \caption{\textbf{Dataset description.} Our dataset contains 63 cases (dementia and MCI) and 63 healthy controls (HC). Acronyms \emph{F} and \emph{M} stand for female and male, respectively. The sum of the gender columns does not correspond to the total number of subjects (126) because 1 of these is undisclosed. The values mentioned within brackets are the standard deviations. The audio lengths are shown in seconds under `Len' and the signal-to-noise ratios (SNR) are shown in dB. In general, audio has very little noise and the long-term prompts are shorter for dementia subjects, but are almost equal between MCI and HC. }
    \label{table:dataset}
    \begin{tabular}{c | c | c | c | c c | c c c c c c c c}
    \hline
    \hline
    \multirow{2}{*}{\textbf{Group}} & \multirow{2}{*}{\textbf{Diagnosis}} & \textbf{Number of} & \multirow{2}{*}{\textbf{Age}} & \multicolumn{2}{c|}{\textbf{Gender}} & \multicolumn{2}{c}{\textbf{Short-term}} & \multicolumn{2}{c}{\textbf{Long-term}} & \multicolumn{2}{c}{\textbf{Semantic-fluency}} & \multicolumn{2}{c}{\textbf{Picture description}}\\ 
    \cline{5-14}
     & & \textbf{Subjects} &  & \textbf{M} & \textbf{F} & \textbf{Len} & \textbf{SNR} & \textbf{Len} & \textbf{SNR} & \textbf{Len} & \textbf{SNR} & \textbf{Len} & \textbf{SNR} \\\hline
     \hline
    \multirow{4}{*}{Case} & \multirow{2}{*}{Dementia}  & \multirow{2}{*}{12} &  75.83   & \multirow{2}{*}{6}  & \multirow{2}{*}{6} & 40.47 & -84.31 & 30.53 & -83.2 & 60.06 & -90.56 & 83.74 & -85.51 \\
     & & & (9.25) & & & (33.79) & (9.23) & (28.07) & (14.4) & (0.13) & (10.97) & (49.36) & (7.92) \\
    \cline{2-14}
     & \multirow{2}{*}{MCI} & \multirow{2}{*}{51} &  68.41 & \multirow{2}{*}{31}  &  \multirow{2}{*}{20} & 39.12 & -82.74 & 41.32 & -82.33 & 60.2 & -85.14 & 76.66 & -88.28 \\ 
      & & & (10.11) & &  & (30.44) & (10.88) & (28.53) & (9.97) & (0.31) & (10.08) & (58.96) & (13.08) \\ \hline
    \multirow{2}{*}{Control} & \multirow{2}{*}{HC} & \multirow{2}{*}{63} & 59.19   & \multirow{2}{*}{28}  & \multirow{2}{*}{34} & 41.12 & -83.6 & 39.51 & -83.62 & 60.18 & -89.76 & 70.19 & -85.17 \\ 
     & & & (15.89) & & & (31.48) & (15.11) & (34.73) & (13.06) & (0.29) & (14.76) & (35.33) & (13.18) \\ \hline
    \multirow{2}{*}{Total}  & \multirow{2}{*}{All}  & \multirow{2}{*}{126} &  64.51 & \multirow{2}{*}{65} & \multirow{2}{*}{60} & 40.25 & -83.32 & 39.39 & -83.06 & 60.18 & -87.96 & 74.10 & -86.46 \\ 
     & & & (14.43) &  &  & (31.3) & (13.06) & (31.88) & (12.06) & (0.29) & (12.92) & (47.77) & (12.82) \\ \hline
     \hline
\end{tabular}
\end{table*}

\vspace{-3pt}

\begin{figure*}[t]
  \centering 
  \includegraphics[width=\linewidth]{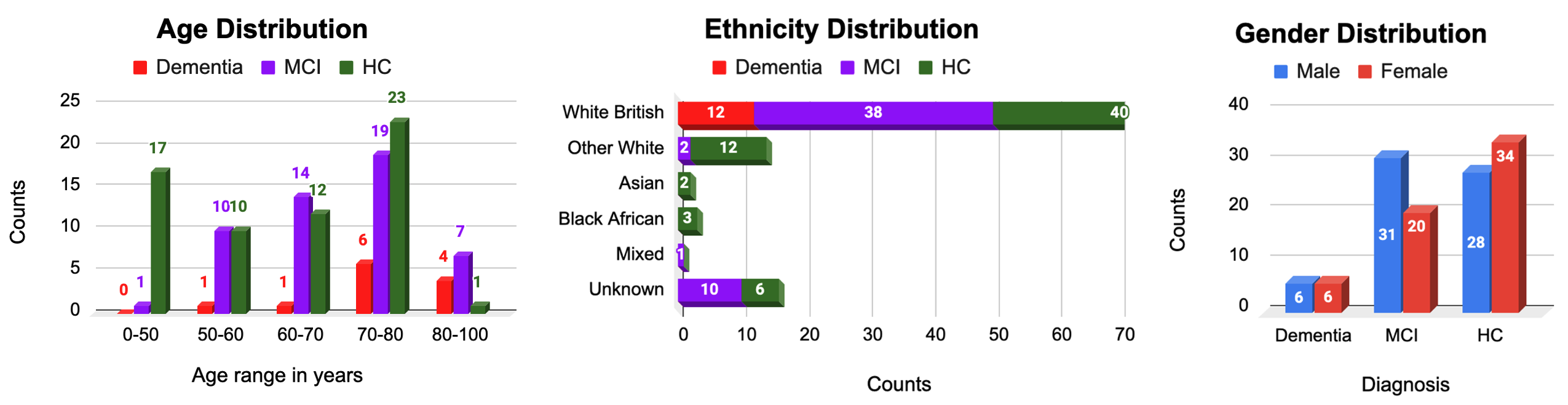}
  \caption{\textbf{Distribution of the demographic information (age, ethnicity and gender) of all 126 subjects used in this study.} The age distribution shows the trend of having more younger people as healthy. The ethnicity distribution shows that even though a small number of subjects represent other ethnicities, most of the participants are white British. Mixed ethnicity is noted for mixed white and black. The gender distribution is almost equally balanced across the groups.}
  \label{fig:demo_dist}
\end{figure*}

\vspace{-3pt}


\section{Previous work}
\label{sec:previouswork}

Analysis of speech and conversation using machine learning techniques has shown promise in the past. 
High accuracies of 74.65\% and 84.51\% to detect AD have been achieved using both acoustic and linguistic features, respectively, from spontaneous speech \cite{pan2021using}. 
In recent years, an increasing number of studies, such as those based on the DementiaBank dataset \cite{becker1994natural} and derivations of it like the ADReSS \cite{luz2020alzheimer} and ADReSSo \cite{luz2021detecting} challenges, worked on picture description data of the \mbox{well-known} cookie theft (CT) picture. 
ADReSS presents challenges of successfully identifying potential AD patients using audio recordings and manual transcriptions, whereas ADReSSo only provided researchers with the audio recordings. 
A special compact set of features from standalone audio helped to achieve an accuracy of 87.6\% \cite{kumar2022dementia}.
Automatic systems like \cite{syed2021tackling, edwards2020multiscale} performed very well by achieving accuracies up to around 90\% using both audio and text (manual transcription) features. 
These public datasets have mostly supported studies on the binary distinction between HC and AD with very little work on MCI. 
One exception has been the work of Mirheidari \etal{} \cite{mirheidari2023identifying}, which used BERT-based features combined with acoustic and textual features to train a classifier and achieve an $F_1$-score of 81.2\% on a dataset of 50 MCI vs.\ 50 HC participants. 
Amini \etal{} \cite{amini2022automated} used \albert{} and \bert{} on a dataset of interviews from three groups of HC (410), MCI (387), and dementia (287) subjects to achieve
an accuracy range between 62\% to 69\% for identifying MCI from HC. 
However, these datasets were small and lacked multiple diagnostic classes, as well as wider ranges of demographics including ethnicity information and rich clinical metadata.

The potential of using automatic speech analysis to monitor the progression of signs of cognitive decline is further confirmed when digital speech assessments completed by older adults over a six-month period were grouped by Montreal Cognitive Assessment (MoCA) scores and adults with higher MoCA scores showed better performance in information richness, language coherence and word-finding abilities. In contrast, MCI and AD adults demonstrated a more rapid decline \cite{robin2021using}. 
As people with early cognitive decline are often not followed closely enough, automatic tools like \cognospeak{} may help monitor longitudinal tracking.


\section{Experimental setup}
\label{section:expsetup}
\subsection{\cognospeak{} data collection}

The participants' data is collected using the \cognospeak{} system, an innovative online AI tool designed to automatically identify potential indicators of cognitive decline. The assessment involves the acquisition of audio and video recordings from participants who are asked to answer a set of questions and complete a series of conventional cognitive tasks by a virtual agent. 
The participants can choose one of the four avatars, specially designed to represent various ethnic and age groups to make the participants feel comfortable speaking to the virtual agent (Figure \ref{fig:cognospeak}). 
The virtual agent prompts have been crafted with input from clinicians and computational linguists to elicit diverse speech patterns and assess speech properties that may be impacted by cognitive impairments. 
For example, the questions include memory recall that aims to examine short or long-term memory \cite{ashford2008screening, lorentz2002brief}, speech fluency, cognitive functioning \cite{cipriani2020daily}, picture description \cite{mueller2018connected} and
reading a 9-sentence or \mbox{129-word} long paragraph \cite{noble2000oral}. 

The answers to these speech elicitation prompts are transcribed automatically to enrich the data with multiple modalities, allowing acoustic and linguistic analysis. All the interactions are also manually transcribed to aid the training of the automatic speech recognition models. In addition, some participants are required to complete a comprehensive suite of questionnaires, such as the MoCA, Multicultural Cognitive Examination (MCE), and Rowland Universal Dementia Assessment Scale (RUDAS), which are intended to evaluate cognitive impairments attributable to dementia. Mental health assessments, including the Patient Health Questionnaire (\mbox{PHQ-9}) and Generalised Anxiety Disorder Assessment (GAD-7), are administered to evaluate the psychological \mbox{well-being} of participants and to allow for the exploration of the confounding effects of cognitive decline and mood on speech parameters. In addition, to achieve cognitive diagnostic labels, participants recruited through the NHS undergo the usual \mbox{gold-standard} diagnosis and participants that are recruited through other routes are asked to complete the Cognitron assessment \cite{brooker2020flame}. To protect privacy, participants are oriented on tool usage and briefed on each question by the virtual agent. Recruitment of participants takes place through a variety of channels, including healthcare referrals, charitable initiatives, and \mbox{word-of-mouth}. All subjects participate voluntarily and sign informed consent items through \cognospeak{}. The data collection is performed under the ethical requirements of the
institutions. 

\cognospeak{} has been developed with extensive user feedback throughout to ensure usefulness and usability. Users from diverse ethnic backgrounds have helped with the \mbox{co-design} of the virtual agents where participants have a choice of four different virtual agents. 
Feedback indicates that users find the system intuitive and easy to use as less than 6\% of users require help to complete the assessment, and only 3.5\% of users dislike using the system. 
In addition, nearly 96\% express positive or neutral views regarding the virtual agent incorporated into the system.

\begin{figure*}[t]
  \centering 
  \includegraphics[width=\linewidth]{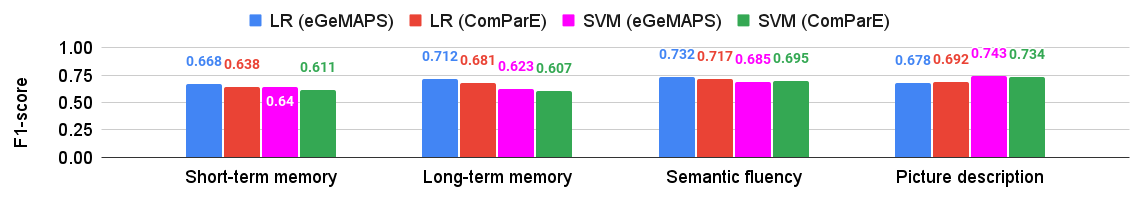}
  \includegraphics[width=\linewidth]{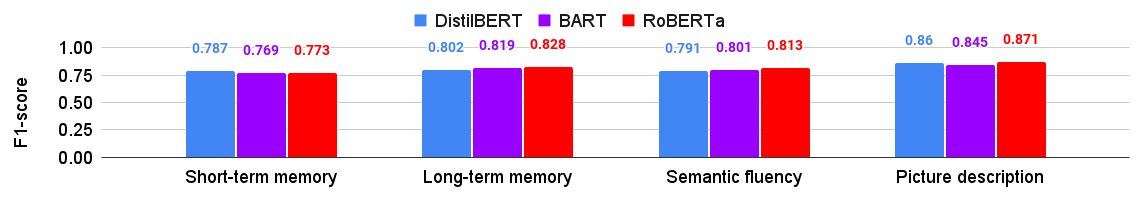}
  \caption{\textbf{Classification based on individual task.} The $F_1$-scores are shown at the top and bottom half while using the acoustic and linguistic features for each classifier respectively. The performance is better when using the linguistic features.}
  \label{fig:barcharts_models}
\end{figure*}

\begin{table*}[t!]
	\setlength{\tabcolsep}{10pt} 
        \centering
        \caption{\textbf{Overall classification performance.} The precision, recall and $F_1$-scores are shown for both acoustic and linguistic features. Again, the foundation models using linguistic features outperform the standard classifiers using acoustic features.}
        \begin{tabular}{cccccc}
        \hline
        \hline
        Experimental Setup & Features & Classifiers & Precision & Recall & $F_1$-score \\ \hline
        \hline
        \multicolumn{1}{c|}{\multirow{4}{*}{Acoustic Features + Standard Classifiers}}    & \multicolumn{1}{c|}{\multirow{2}{*}{\egemaps{}}} & LR       & $0.682\pm0.9$  & $0.678\pm0.9$ & $0.665\pm0.9$\\
        \multicolumn{1}{c|}{}  & \multicolumn{1}{c|}{} & SVM & $0.691\pm0.8$  & $0.681\pm0.8$ & $0.684\pm0.8$\\ \cline{2-6} 
        \multicolumn{1}{c|}{} & \multicolumn{1}{c|}{\multirow{2}{*}{\compare{}}} & LR & $0.691\pm0.6$  & $0.720\pm0.6$ & $0.702\pm0.6$\\
        \multicolumn{1}{c|}{}  & \multicolumn{1}{c|}{} & SVM  & $0.721\pm0.7$  & $0.724\pm0.7$ & $0.723\pm0.7$\\ \hline
        
        \multicolumn{1}{c|}{\multirow{3}{*}{Linguistic Features + Foundation Models}} & \multicolumn{1}{c|}{\multirow{3}{*}{ASR}}     & \distilbert{} &  $\mathbf{0.875\pm0.2}$     &  $\mathbf{0.871\pm0.2}$   & $\mathbf{0.873\pm0.2}$   \\
        \multicolumn{1}{c|}{}  & \multicolumn{1}{c|}{}                         & \bart{}     &  $0.834\pm0.4$     &  $0.858\pm0.4$   &  $0.847\pm0.4$ \\
        \multicolumn{1}{c|}{}  & \multicolumn{1}{c|}{}                         & \roberta{}  &   $0.881\pm0.3$    &   $0.864\pm0.3$  &  $0.868\pm0.3$  \\ 
        \hline
        \hline
\end{tabular}
\label{table:results}
\end{table*}


\subsection{Dataset description}

Data collection is ongoing, and to date, more than 1600 participants have completed the assessment and 12.63\% of them are from UK ethnic minorities for which we collect a range of language-related metadata enabling us to explore robustness to language background and potential bias issues.  
Of these, manual transcripts are so far available for a subset of the participants, specifically 126 individuals for the experiments conducted in this study. 
To explore the individual merits of the various elicitation tasks, we have conducted experiments on four sets of tasks, namely the short-term memory question, the long-term memory question, the semantic fluency task and the CT picture description task. 
The short and long-term memory probing questions were `What did you do over the weekend? Please give as much detail as possible' and `Please could you tell me about the school you went to and how old you were when you left?'
The subject was asked to name as many animals as possible within a minute for the semantic fluency task and then invited to describe the CT picture.
The total length of recordings for this cohort is 7 hours, 29 minutes and 13 seconds, which includes 12 individuals diagnosed with dementia, 51 diagnosed with MCI, and 63 participants identified as healthy controls, i.e., having no cognitive health issues. 
Table~\ref{table:dataset} shows the prompt lengths and signal-to-noise ratio (SNR) along with its standard deviation for each group as well. 
Audio in general contains very little noise as the lowest SNR has been -83.06$\pm$12.06 dB for long-term memory tasks as shown in Table \ref{table:dataset}. 

Table~\ref{table:dataset} shows the demographic information in terms of the number of subjects, age, and gender and Figure \ref{fig:demo_dist} shows the distribution of age, ethnicity and gender. 
They demonstrate that the gender distribution is almost equally balanced among both cases (dementia and MCI) and healthy controls. 
However, as the data has been collected mainly in England so far, the majority of the participants are white British. Figure \ref{fig:demo_dist} also shows that all dementia and MCI subjects are white, except 10 MCI subjects who did not disclose their ethnic information. 
According to a two-tailed $t$-test ($p < 0.001$), no statistically significant difference in gender distribution was identified, however, there is a statistically significant difference between the age distribution of patients, who were diagnosed with dementia and MCI and have an average age of 69.83$\pm$10.37 years, and healthy controls. 
This observation aligns with the consensus in neuroscience that cognitive decline, often attributed to structural and functional changes within the brain, predominantly occurs with aging~\cite{murman2015impact}.

\subsection{Feature extraction and classification}\label{subsec:feat_class}

The audio of the selected four tasks was preprocessed by forced alignment and segmentation before extracting both acoustic and linguistic features, as shown in Figure \ref{fig:cognospeak}. 
Acoustic features include \egemaps{} and \compare{} 2016 from OpenSmile 3.0 \cite{eyben2015geneva, eyben2010opensmile}. 
Foundation models are proven useful in sentiment, language and text analysis in recent times\cite{zhang2023text}, and here we explore their abilities in detecting cognitive impairments \cite{bommasani2021opportunities}. 
We have applied three such models which are \bart{} \cite{lewis2019bart}, \roberta{}
\cite{liu2019roberta, matovsevic2022accurate}
and
\distilbert
\cite{sanh2019distilbert, nambiar2022comparative}. 
Initial experiments using different seeds on these models exhibited no significant improvements.
The experiments were performed over $k$-fold cross-validation ($k=5$) with the same data split with strictly no overlap between folds to make rigorous comparisons between approaches and make the best use of the relatively small amount of data in each class.


\section{Results}
\label{sec:results}

\subsection{Classification on individual tests}
Figure \ref{fig:barcharts_models} shows the
$F_1$-scores achieved while using individual tasks for classification.
The top half shows the results using the acoustic features while the bottom half shows the results for linguistics features. 
\subsubsection{Acoustic features}
For the short-term memory task, a logistic regression (LR) classifier using \egemaps{} has performed the best, followed by the support vector machine (SVM), as they produced the $F_1$-scores of 0.668 and 0.64 respectively. Here, \egemaps{} features performed better than the \compare{} features.
The same pattern is also visible for long-term memory tasks as the LR using \egemaps{} again outperformed the others. It achieved an $F_1$-score of 0.712, and the next best performance was also achieved by LR, but this time using \compare{} features. 
Figure \ref{fig:barcharts_models} also shows that the variations are quite high between performances across the classifiers for using long-term memory tasks. 
An \mbox{$F_1$-score} of 0.732 has also been achieved from an LR for the semantic fluency task using the \egemaps{} features. \compare{} features performed close to equally well by producing the $F_1$-score of 0.717. 
SVM has produced the highest $F_1$-score of 0.743 and 0.734 while using the \egemaps{} and \compare{} features extracted from the picture description respectively. 
A lightly lower \mbox{$F_1$-scores} of 0.678 and 0.692 have been achieved from those features while using the LR classifier. 
Figure \ref{fig:barcharts_models} demonstrates that LR has generally performed better than SVM while using acoustic features for short and long-term memory tasks, and semantic memory tasks individually.

\subsubsection{Linguistic features}
The bottom half of Figure \ref{fig:barcharts_models} shows the performances of the linguistic-based foundation models. As expected, overall, these models outperform the more conventional acoustic-based classifiers.
\distilbert{} has performed the best while using short-term memory tasks by achieving the $F_1$-score of 0.787, which was closely followed by \roberta{} as it produced the $F_1$-score of 0.773.
\bart{} has performed similarly as its $F_1$-score is 0.769. 
However, \roberta{} has outperformed all others by achieving an \mbox{$F_1$-score} of 0.828 while using the long-term memory tasks as \bart{} and \distilbert{} have produced the $F_1$-scores of 0.819 and 0.802 respectively. 
For semantic fluency tests, \roberta{} has achieved the \mbox{$F_1$-score} of 0.813, thus outperforming others by a small margin, as \bart{} and \distilbert{} have produced the scores of 0.801 and 0.791 respectively. 
Finally, \roberta{} has performed the best by achieving an \mbox{$F_1$-score} of 0.871 while using the features extracted from the picture description tasks. \distilbert{} and \bart{} have also performed almost equally well by producing $F_1$-scores of 0.86 and 0.845. 

\vspace{-5pt}

\subsection{Overall classification performance}
The classification was also performed by considering all four tasks and applying a majority voting algorithm to predict the final label of a test subject. 
The precision, recall and $F_1$-scores along with their standard deviations ($\sigma$) across the cross-validation folds are reported as the overall performances in Table \ref{table:results}.
The LR has produced an $F_1$-score of 0.665, which is slightly less than the $F_1$-score of 0.684 generated by the SVM using \egemaps{} features as overall performance. 
These performances are better while using the overall \compare{} features as $F_1$-scores of 0.702 and 0.723 were achieved from LR and SVM. 
The overall performances of the foundation models using linguistic features are statistically significantly higher than classifiers using acoustic features. 
Here, \distilbert{} have outperformed all others in overall performance by achieving the highest $F_1$-score of 0.873 with precision and recall values of 0.875 and 0.871 respectively with 
a $\sigma$ of 0.2.
\roberta{} has achieved an almost equal performance by generating \mbox{$F_1$-score} of 0.868 with the $\sigma$ of 0.3 and \bart{} has achieved a slightly lower $F_1$-score of 0.847 with the $\sigma$ of 0.4.

\vspace{-5pt}

\begin{figure}[ht]
  \centering 
  \includegraphics[width=0.99\linewidth]{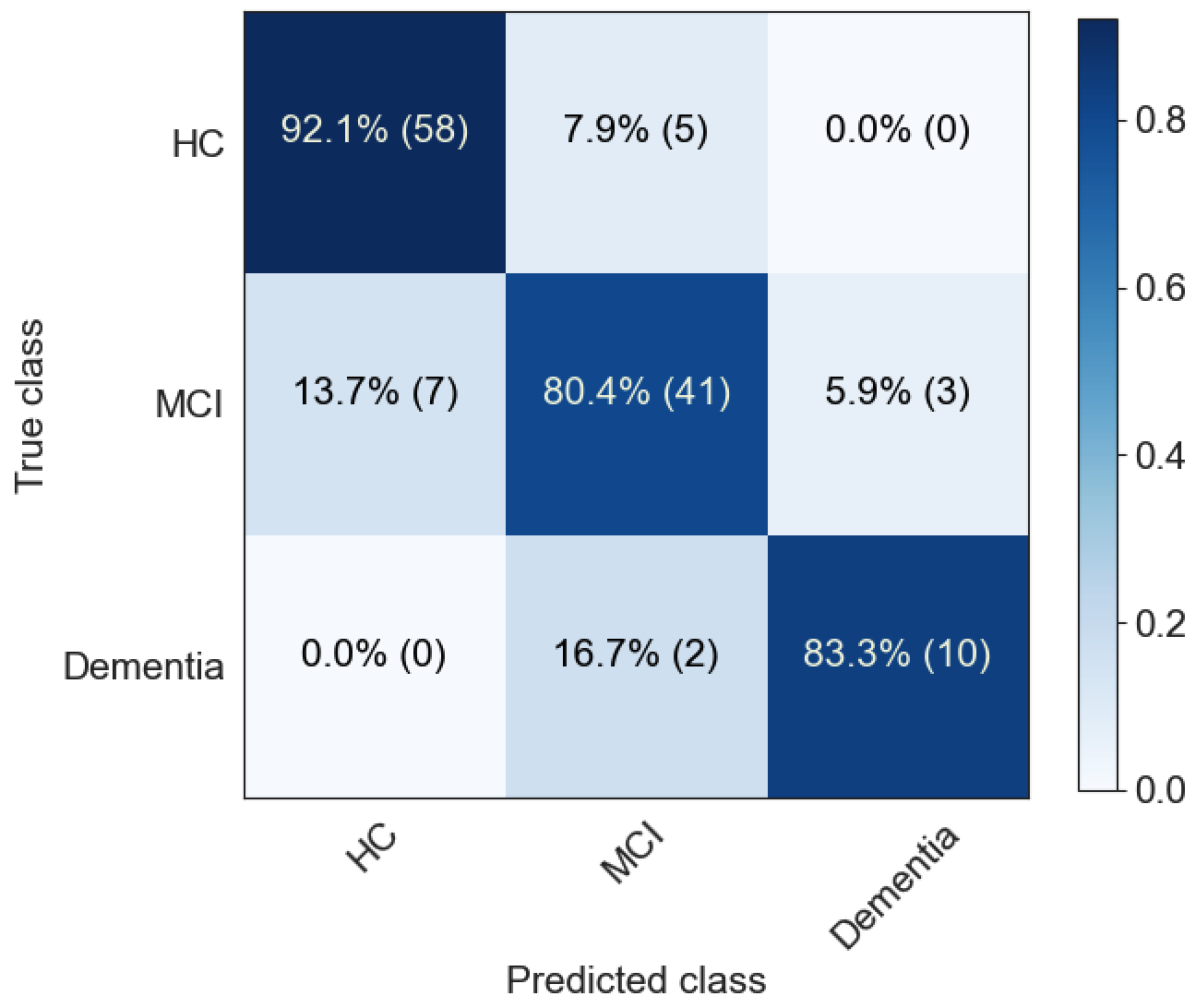} 
  \caption{ \textbf{Confusion matrix.} The best-performing \distilbert{} model, shown in Table \ref{table:results}, was applied to all test folds of cross-validation. Detecting MCI was the most challenging.}  
  \label{fig:cm_roberta_model}
\end{figure}

\vspace{-5pt}


\section{Discussion}
\label{sec:discussion}  
The results show that, in general, LR has performed better than SVM, and \egemaps{} have been the features of choice while using the acoustic features for classification. 
Results also provide evidence of using semantic fluency tests, long-term memory tests and picture description tasks as the primary means of detecting early signs of cognitive decline using acoustic features as they produced the highest $F_1$-scores of 0.743, 0.732 and 0.712 respectively. 
While using the overall acoustic features, \compare{} has been the feature of choice as both LR and SVM performed better than \egemaps{}. 
However, linguistic features have outperformed the acoustic features by a large margin for both individual and overall performance. 
Table \ref{table:results} also demonstrates that the $\sigma$ among the cross-validation folds are lower for the linguistic models as well, indicating better generalisation over a diverse dataset. 
These observations also support our previous findings where a complex, pre-trained and fine-tuned architecture performed better than the standard classifiers \cite{pahar2022covid}. 

Figure~\ref{fig:cm_roberta_model} presents the confusion matrix which is generated by applying the \mbox{best-performed} \distilbert{} model, which achieved the highest $F_1$-score of 0.873 using all four tasks 
(Table \ref{table:results}), 
to each cross-validation test fold and the actual and predicted labels are summed up. 
It exhibits good overall performance as subjects with no cognitive impairment were predicted with 92.1\% accuracy, followed by 83.3\% and 80.4\% accuracy for MCI and dementia. 
This can also be translated into a two-class confusion matrix, which shows a specificity of 92.1\% and sensitivity of 81\% for detecting either dementia or MCI. 
As cognitive declines of those suffering from MCI are hard to notice, it is more challenging for the foundation models to predict accurately and hence a very promising result, despite the skewness present in the age and ethnicity distribution in our data as shown in Figure \ref{fig:demo_dist}. 



\section{Conclusion}
\label{sec:conclusion}  
We have presented an automatic remote assessment tool for detecting early signs of cognitive decline using real-world conversational speech. 
Speaking requires significant cognitive resources, including memory, language, and attention thus carrying the early signs of cognitive impairment. 
\cognospeak{} engages participants in a conversation using a virtual agent and asks memory-probing questions alongside administering more conventional cognitive tests. 
To provide an initial set of results that can be used as comparisons based on the data collected by \cognospeak{} in future work, a subset of 126 subjects, of whom there are 12 dementia, 51 MCI and 63 healthy controls, was prepared. Both acoustic and linguistic features were extracted from short-term memory, long-term memory, semantic fluency and picture description tasks.
The linguistic features extracted from the combination of these four tasks with \distilbert{} model have performed the best by producing the highest \mbox{$F_1$-score} of 0.873. 
The corresponding confusion matrix shows that predicting healthy controls has been the most successful, followed by dementia and MCI. 
\cognospeak{} provides a \mbox{low-cost}, repeatable, non-invasive and less stressful alternative to the current cognitive assessment methods to detect early signs of cognitive decline.

As for our immediate future work, rigorous classifier training and fine-tuning will be carried out to improve the classification performance on a larger corpus, which will contain less skewed demographic information and will be extended to multi-class classification. 
The performance of \cognospeak{} as an automatic remote long-term monitoring tool will also be investigated using the follow-up patients alongside their MoCA, MCE, RUDAS, PHQ-9 and GAD-7 scores. 
Finally, the entire multimodal data corpus along with the rich metadata will be released to the research community soon and a small subset of this data has already been shared with more than 80 research groups around the world as a part of 
the 
``The Prediction and Recognition Of Cognitive declinE through Spontaneous Speech (PROCESS)'' 
signal processing grand challenge in ICASSP 2025.


\bibliographystyle{IEEEtran}
\bibliography{reference}

\end{document}